# Study of the possibility of eliminating the Gibbs paradox within the framework of classical thermodynamics *

## V. Ihnatovych

Department of Philosophy, National Technical University of Ukraine "Kyiv Polytechnic Institute",
Kyiv, Ukraine
e-mail: V.Ihnatovych@kpi.ua

### Abstract

The formulas for the entropy of ideal gases mixture and the entropy change in mixing of ideal gases on the basis of the third law of thermodynamics were obtained. It is shown that when using these formulas, the Gibbs paradox within the framework of classical thermodynamics does not arise.

### 1. Introduction

The Gibbs paradox arose in the theoretical considering a question of a change in the entropy in mixing of two ideal gases. It consists in that the value of the change of entropy during the mixing of two ideal gases (entropy of mixing), with the same initial temperature and pressure does not dependent on the properties of the mixed gases (until they are different), and unevenly becomes zero in the transition from mixing of different gases to mixing of identical gases. A jump of the entropy of mixing is paradoxical, when parameter of gases difference goes to zero without a jump [1-10].

The search of this paradox explanations lasts for more than 100 years and is described in detail in the monograph [10], where about fifty original explanations for this paradox are given and a conclusion that there is no generally accepted explanation is made.

In considering this paradox, the various authors discussed, in particular (and did not come to common opinion), as to whether the conclusion of the jump of entropy of mixing is connected with the assumption of the existence of discrete differences between the parameters of the mixed gases or not, and can the conclusion of this jump be eliminated, if we accept the premise of the possibility of a smooth transition from one gas to another [1-10].

**In our opinion, many questions about this paradox are not solved because in considering this paradox its logical and mathematical aspects were not taken into account: the point that the conclusion about the paradoxical jump was made in certain discourses and regards to mathematical behavior of the functions of many variables – the entropy of ideal gases mixing, for which we can derive the formula was not taken into account.**

If these circumstances are taken into account, it must be said that the characteristics of peculiarities of behavior of the function for which there is a formula, are determined by the peculiarities of this formula and the peculiarities of behavior of variables and parameters that are included in the formula. The formula for the entropy of mixing is obtained by the development on the basis of certain initial formulas. It is understood that the peculiarities of behavior of the entropy of mixing are caused by the peculiarities of the formulas on which basis the formula for the entropy of mixing is developed. If we develop a formula for the entropy of mixing, using other initial formulas, we can get the other peculiarities of the behavior of the entropy of mixing.

In this paper the author intends to show that the corresponding formulas can be obtained in classical thermodynamics and thus eliminate the Gibbs paradox in classical thermodynamics.

## 2. Preliminaries

In the papers concerning the Gibbs paradox, the following formulas are used as the initial ones:

$$\Delta S_m = S_m - S_g, \tag{1}$$

$$S_g = \sum S_j, \tag{2}$$

$$S_m = \sum S_i, \tag{3}$$

$$S_i = n_i(c_{pi} \ln T_i - R \ln p_i + S_{0pi}) \tag{4}$$

where $\Delta S_m$ – the entropy of mixing, $S_m$ – the entropy of the mixture, $S_g$ – the entropy of the system consisting of subsystems, separated by impermeable partitions, $S_j$ – the entropy of $j$-th subsystem, $S_i$ – the entropy of $i$-th ideal gas – a mixture component, $n_i$ – the number of moles of the $i$-th gas, $c_{pi}$ – its molar thermal capacity at constant pressure, $T_i$ – thermodynamic temperature, $S_{0pi}$ – integration constant, which depends on the nature of the gas and does not depend on $n, c, V, p, T$.

Formulas (2) – (4) express the absolute value of the entropy of thermodynamic systems. In classical thermodynamics, the absolute value of the entropy is determined on the basis of the second and third laws of thermodynamics.

According to the second law of thermodynamics,

$$dS = \frac{\delta Q}{T}, \tag{5}$$

where $\delta Q$ – the amount of heat absorbed by the system in elementary equilibrium process.

According to the third law of thermodynamics, the entropy of a pure substance – an ideal crystal – when $T = 0$ equals to zero.

Accordingly, as to the second and third laws of thermodynamics, the absolute value of the entropy of any pure substance, which when $T = 0$ is an ideal crystal, equals to:

$$S_i = \int_0^T \frac{\delta Q_p}{T} = \int_0^T \frac{dH}{T} + \sum \frac{\Delta H_{kph}}{T_{kph}} = \int_0^T \frac{C_{pi}}{T} dT + \sum \frac{\Delta H_{kph}}{T_{kph}}. \tag{6}$$

where $Q_p$ – the heat that is absorbed by the system at constant pressure; $H$ – the enthalpy of the system; $\Delta H_{kph}$ – the enthalpy change in $k$-th phase transition, $T_{kph}$ – the temperature of $k$-th phase transition, $C_{pi}$ – the thermal capacities of substance at constant pressure.

The formula (6) can be used to determine the entropy of substance, which is an ideal gas at a given pressure $p$ and temperature $T \geq T_1$. Of course, when the temperature lows at constant pressure such substance becomes not ideal gas (real gas), then – liquid and solid.

At $T \geq T_1$ the entropy of such substance can be determined by the formulas (4) and (6). Accordingly, if

$$S_0 = \int_0^T \frac{C_{pi}}{T} dT + \sum \frac{\Delta H_{kph}}{T_{kph}} - C_p \ln T_1 + R \ln p_1.$$

then the formula (6) agrees with the formula (4).

The entropy of the multicomponent system, which is at temperature $T$ of a mixture of ideal gases, is equal to:

$$S_i = \int_0^T \frac{\delta Q_p}{T} = \int_0^T \frac{dH_m}{T} + \sum \frac{\Delta H_{kmph}}{T_{kmph}} + S_{m0} = \int_0^T \frac{C_{pm}}{T} dT + \sum \frac{\Delta H_{kmph}}{T_{kmph}} + S_{m0}. \tag{7}$$

where $H_m$ – the enthalpy of the system; $C_{pm}$ – the thermal capacities of the system at constant pressure, $\Delta H_{kmph}$ – the enthalpy change in $k$-th phase transition in the system; $T_{kmph}$ – the temperature of $k$-th phase transition in the system, $S_{m0}$ – the entropy of the system when $T = 0$.

In the formula (7) the fact that the entropy of the multicomponent system at $T = 0$ can differ from zero is taken into account.

For a system, which is composed of gases with equal temperature, separated by partitions,

$$Q = \Sigma Q_i.  \quad (8)$$

From (5) and (8) follows (2).

## 3. Definition of the entropy of ideal gases mixing using the third law of thermodynamics

In order not to be distracted by minor details, we consider the case of mixing of two ideal gases with equal initial temperatures and pressures, previously separated by an impermeable partition. We also assume that the system contains one mole of each gas.

For this case of (1), (6) – (8) follows:

$$\Delta S_m = \int_0^T \frac{C_{pm} - (c_{p1} + c_{p2})}{T} dT + \sum \frac{\Delta H_{kmph}}{T_{kmph}} - \left( \sum \frac{\Delta H_{kph1}}{T_{kph1}} + \sum \frac{\Delta H_{kph2}}{T_{kph2}} \right) + S_{m0}. \quad (9)$$

If we use the formulas (2) – (4), for a mixture of different gases we obtain

$$\Delta S_m = 2R \ln 2, \quad (10)$$

and for a mixture of identical gases

$$\Delta S_m = 0. \quad (11)$$

According to (9), the value of the entropy of ideal gases mixing (at $T \geq T_1$) depends on the difference in the behavior of functions $C_{pm}$, $c_{p1} + c_{p2}$, $\frac{\Delta H_{kmph}}{T_{kmph}}$, $\left( \sum \frac{\Delta H_{kph1}}{T_{kph1}} + \sum \frac{\Delta H_{kph2}}{T_{kph2}} \right)$ in the temperature range $0 \ldots T_1$, which complicated depends on the properties the gases, as at low temperatures, gases and their mixtures are not ideal and at certain temperatures are moving to a condensed state. It is known that the closer the mixture components are by their properties, the closer are the values of functions $C_{pm}$ and $c_{p1} + c_{p2}$, $\frac{\Delta H_{kmph}}{T_{kmph}}$ and $\left( \sum \frac{\Delta H_{kph1}}{T_{kph1}} + \sum \frac{\Delta H_{kph2}}{T_{kph2}} \right)$ and according to (9), the closer to zero is the value of $\Delta S$. Consequently, in determining the entropy of gases and mixtures on the basis of the second and third laws of thermodynamics there is no question of the independence of the entropy of mixing of the kind of gases, which, according S.D. Khaitun [1, p. 24], is an integral part of the Gibbs paradox. The mixtures of optical isomers are ideal. Therefore, for different gases – optical isomers – the value determined by the formula (9) is equal to zero. When determining the entropy according to the equation (9) there is also a jump of entropy of mixing in the transition from the mixing of different to the mixing of identical gases.

If the value $\Delta S_c$ has any peculiarities of behavior when approaching the properties of mixed gases, then, according to (9), they may be due only to the peculiarities of behavior of values $C_{pm}$, $c_{pi}$, $\frac{\Delta H_{kmph}}{T_{kmph}}$, $\frac{\Delta H_{kphi}}{T_{kphi}}$. In particular, the jump of the entropy of mixing may take place only in the event when any the indicated values changes unevenly. Respectively, the entropy of mixing, determined using the third law of thermodynamics, cannot have any peculiarities of the behavior, not connected with the peculiarities of behavior of heat, enthalpy or internal energy of ideal gases.

Thus, if we start from the third law of thermodynamics, it is impossible to obtain conclusions regarding the behavior of the entropy of ideal gases mixing that make up the content of the Gibbs paradox.

Due to the fact that the results, obtained using the third law of thermodynamics, contradict to the results obtained on the basis of formulas (2) – (4), at least one of the formulas (2) – (4) contradicts the third law of thermodynamics.

It is shown above that the formula (6) does not contradict to the formula (4), and that the formula (2) can be used to determine the entropy of ideal gases mixing using the third law of thermodynamics.

Therefore, the differences of behavior of the value $\Delta S_m$, obtained on the basis of the formulas (2) – (4) and the value $\Delta S_m$, obtained on the basis of the formulas (5) – (8), are conditioned by the differences of behavior of values $S_m$, determined by the formulas (3) and (7). Therefore, we can assume that the formula (3), which expresses the Gibbs theorem of entropy of ideal gases mixing, cannot be reconciled with the third law of thermodynamics.

Let us note that in the classical thermodynamics, Gibbs theorem is proven on the basis of Dalton law about the pressure of ideal gases mixture (see, for example [1 – 3, 8]). It is easy to make sure that we can strictly get only the formula for the entropy change when the state of an ideal gas mixture changes from the formula of the entropy of an ideal gas and Dalton law:

$$\Delta S_m = \sum \Delta S_i, \qquad (12)$$

and the formula (3) does not follow it.

## 4. Conclusions

Gibbs paradox in the formulation of the peculiarities of mathematical behavior of the entropy of ideal gases mixing in the transition from mixing of different gases to mixing of identical gases in the classical thermodynamics can be eliminated from this theory, if for definition of the entropy of mixing we use the third law of thermodynamics instead of the Gibbs theorem about the entropy of an ideal gases mixing.

It can be assumed that the Gibbs theorem about the entropy of ideal gases mixing contradicts to formulas, based on the second and third laws of thermodynamics.